\documentclass[preprint,amsfonts,amssymb,amsmath,%
tightenlines,nofootinbib,showpacs,noshowkeys]{revtex4}

\newcommand{\cD}{\mathcal{D}}

\newcommand{\cN}{\mathcal{N}}
\newcommand{\cP}{\mathcal{P}}
\newcommand{\cT}{\mathcal{T}}
\newcommand{\cV}{\mathcal{V}}
\newcommand{\cW}{\mathcal{W}}
\newcommand{\PT}{\mathcal{PT}}
\newcommand{\tH}{\tilde{H}}

\newcommand{\tcV}{\tilde{\cV}}
\newcommand{\bA}{\mathbf{A}}
\newcommand{\bH}{\mathbf{H}}
\newcommand{\bQ}{\mathbf{Q}}
\newcommand{\xb}{\bar{x}}
\newcommand{\bbR}{\mathbb{R}}

\newcommand{\rmi}{\mathrm{i}}
\newcommand{\rme}{\mathrm{e}}
\newcommand{\rmd}{\mathrm{d}}
\newcommand{\rmf}{\mathrm{f}}
\newcommand{\fsl}{\mathfrak{sl}}

\begin{document}



%
%

\title{Nonlinear Pseudo-Supersymmetry in the Framework of\\
$\cN$-fold Supersymmetry}
\author{Artemio Gonz\'{a}lez-L\'{o}pez}
\email{artemio@fis.ucm.es}
\affiliation{Departamento de F\'{\i}sica Te\'{o}rica II,
 Facultad de Ciencias F\'{\i}sicas,
 Universidad Complutense, 28040 Madrid, Spain}
\author{Toshiaki Tanaka}
\email{ttanaka@mail.tku.edu.tw}
\affiliation{Department of Physics, Tamkang University,
 Tamsui 25137, Taiwan, R.O.C.}


\begin{abstract}

We recall the importance of recognizing the different mathematical
nature of various concepts relating to $\PT$-symmetric quantum
theories. After clarifying the relation between supersymmetry
and pseudo-supersymmetry, we prove generically that nonlinear
pseudo-supersymmetry, recently proposed by Sinha and Roy, is
just a special case of $\cN$-fold supersymmetry. In particular,
we show that all the models constructed by these authors have
type A 2-fold supersymmetry. Furthermore, we prove that
an arbitrary one-body quantum Hamiltonian which admits
two (local) solutions in closed form belongs to type A
2-fold supersymmetry, irrespective of whether or not it
is Hermitian, $\PT$-symmetric, pseudo-Hermitian, and so on.

\end{abstract}


\pacs{03.65.Ge; 02.30.Hq; 11.30.Pb; 11.30.Na}




\maketitle

\section{Introduction}

Since Bender and Boettcher claimed that the reality of the spectrum
of the Hamiltonian $H=p^{2}+x^{2}+\rmi x^{3}$ is due to the underlying
$\PT$ symmetry \cite{BB98a}, there have appeared in the literature
numerous investigations into the spectral properties of various
quantum Hamiltonians with non-real potentials defined on, in general,
a complex contour. See, e.g.,
Refs.~\cite{Sh05,GSZ05,Gh05,Tr05,BQR05,BJR05}
and references cited therein for recent developments.
The rapid progress in this research field, however, has caused
some confusion and several misunderstandings. A typical
example is the relation between $\PT$ symmetry and
pseudo-Hermiticity. The former concept is meaningful at the
operator level, without referring to a vector space on which the
operator in question acts; indeed, it can be defined as the invariance
of the operator under the formal replacements $x\to -x$ and
$\rmi\to -\rmi$. On the other hand, the concept of pseudo-Hermiticity,
mainly developed by Mostafazadeh in the context of $\PT$ symmetry
\cite{Mo02a}, inevitably needs a Hilbert space on which the Hermitian
conjugate is defined. Hence, it makes little sense to discuss, e.g.,
whether or not $\PT$ symmetry is a special case of pseudo-Hermiticity,
without taking into account their different mathematical character.

Recently, we have also found a similar confusion in the paper by Sinha
and Roy \cite{SR05}, where the authors claimed to generalize
the framework of $\cN$-fold supersymmetry to include pseudo-Hermitian
systems. This misunderstanding is apparently inherited from the claim
in Ref.~\cite{Mo02a} that pseudo-supersymmetry is a generalization of
(ordinary) supersymmetry.

Considering the current situation in and around this research field,
we would like to recall in this paper the importance of recognizing
the different mathematical nature of various concepts relating
to $\PT$-symmetric quantum theories.
In particular, we focus on the relation among the nonlinear
pseudo-supersymmetric models in Ref.~\cite{SR05}, the framework of
$\cN$-fold supersymmetry, and higher-order Darboux transformations.

The paper is organized as follows. In the next section, we review
the definition of various concepts characterizing linear differential
operators, which are relevant in $\PT$-symmetric quantum theories,
to avoid ambiguity. Based on these precise definitions, we clarify
the exact relation among various types of supersymmetry in
Section~\ref{sec:sps}. We then proceed to prove that nonlinear
pseudo-supersymmetry automatically implies $\cN$-fold supersymmetry
in Section~\ref{sec:Nsnps}. To make the relation more transparent,
we also show how all the models in Ref.~\cite{SR05} can be explicitly
constructed in the framework of $\cN$-fold supersymmetry. These
findings clearly suggest that there is an overlooked relation between
the higher-order Darboux transformations and $\cN$-fold supersymmetry,
which we discuss in Section~\ref{sec:2DtA2s}. The paper concludes
in Section~\ref{sec:concl} with a short discussion of the main results
obtained in it and some general remarks on the different mathematical
character of the symmetries considered.

\section{Pseudo-{H}ermiticity and $\PT$ Symmetry}
\label{sec:PHPT}

First of all, we would like to review the definition of $\PT$
symmetry \cite{BB98a} and pseudo-Hermiticity first introduced in
Ref.~\cite{Mo02a}. In this paper, we restrict our discussion to
linear operators acting on a linear function space of a single
variable, e.g., $x$, which have generally the following form:
\begin{align}
L=\sum_{n=0}^{\infty}f_{n}(x)\frac{\rmd^{n}}{\rmd x^{n}}.
\label{eq:lop}
\end{align}

We first define the \emph{formal} Hermitian conjugate $L^{\rmf}$ of
the operator \eqref{eq:lop} by
\begin{align}
L^{\rmf}=\sum_{n=0}^{\infty}(-1)^{n}\frac{\rmd^{n}}{\rmd x^{n}}
 f_{n}^{\ast}(x^{\ast}),
\end{align}
where $\ast$ denotes complex conjugate. A linear operator $L$ is
called \emph{formally Hermitian} if $L^{\rmf}=L$. Similarly, the
transposition $L^{t}$ of the operator \eqref{eq:lop} is defined by
\begin{align}
L^{t}=\sum_{n=0}^{\infty}(-1)^{n}\frac{\rmd^{n}}{\rmd x^{n}}
 f_{n}(x),
\end{align}
and $L$ is said to have \emph{transposition symmetry} if
$L^{t}=L$ \cite{AS03}. We note that the formal Hermitian conjugate
and transposition of a product of two linear operators
formally satisfies $(L_{1}L_{2})^{\rmf}=L_{2}^{\rmf}L_{1}^{\rmf}$
and $(L_{1}L_{2})^{t}=L_{2}^{t}L_{1}^{t}$, respectively,
by the above definition.

The spatial reflection $\cP$ and the time reversal $\cT$ of
the operator \eqref{eq:lop} are, respectively, defined by
\begin{align}
\cP L\cP&=\sum_{n=0}^{\infty}(-1)^{n}f_{n}(-x)
 \frac{\rmd^{n}}{\rmd x^{n}},\\
\cT L\cT&=\sum_{n=0}^{\infty}f_{n}^{\ast}(x^{\ast})
 \frac{\rmd^{n}}{\rmd x^{n}},
\end{align}
where we note that $\cP^{2}=\cT^{2}=1$ and $\cP\cT=\cT\cP$.
A linear operator $L$ is said to have \emph{$\PT$ symmetry} if
$\PT L\PT=L$ \cite{BB98a}.

Let $\cV$ be a linear function space, let $\eta$ be an invertible,
formally Hermitian operator on $\cV$, and consider a linear
differential operator $L: \cV\to\cV$ of the form \eqref{eq:lop}. Then,
the \emph{formal} pseudo-Hermitian conjugate $L^{\natural}: \cV\to\cV$
with respect to $\eta$ is defined by $L^{\natural}=\eta^{-1}L^{\rmf}
\eta$. A linear operator \eqref{eq:lop} is called \emph{formally
pseudo-Hermitian} if there exists an invertible, formally Hermitian
operator $\eta$ satisfying $L^{\natural}=L$, or equivalently,
$L^{\rmf}=\eta L\eta^{-1}$. It is evident that formal
pseudo-Hermiticity reduces to formal Hermiticity when $\eta=1$.

The Hermitian conjugate of the linear differential
operator \eqref{eq:lop} acting on a Hilbert space $L^{2}(S)$
($S\subset\bbR$) with the positive definite inner product
$(\phi,\psi)$ defined by
\begin{align}
(\phi,\psi)=\int_{S}\rmd x \,\phi^{\ast}(x)\psi(x)
\label{eq:inner}
\end{align}
is the operator $L^{\dagger}$ satisfying
\begin{align}
(\phi,L^{\dagger}\psi)=(L\phi,\psi),
 \qquad\forall\phi,\psi\in L^{2}(S),
\end{align}
and formally coincides with the formal Hermitian conjugate
$L^{\rmf}$. Then, the operator $L$ is called \emph{Hermitian}
(or\emph{self-adjoint}) if $L^{\dagger}=L$ with respect to the inner
product \eqref{eq:inner}\footnote{Note, in particular, that
$\cD(L)=\cD(L^{\dagger})$, where $\cD$ denotes the domain of the
operator.}. It is evident that a Hermitian operator on $L^{2}(S)$ is
always formally Hermitian. Similarly, the operator $L$ is called
\emph{pseudo-Hermitian} if there exists an invertible, Hermitian
operator $\eta$ satisfying $L^{\dagger}=\eta L\eta^{-1}$ with respect
to the inner product \eqref{eq:inner} \cite{Mo02a}. It is also evident
that a pseudo-Hermitian operator on $L^{2}(S)$ is always formally
pseudo-Hermitian.

The crucial problems in the construction of pseudo-Hermitian theories
are that the eigenvectors of a pseudo-Hermitian operator are not
in general orthogonal with respect to the inner product
\eqref{eq:inner}, and that ascertaining that these eigenstates
span a dense set of the Hilbert space $L^{2}(S)$ is far from
trivial. These facts clearly indicate the difficulty in establishing,
e.g., a resolution of the identity and a spectral theorem for
pseudo-Hermitian operators in terms of orthogonal spectral
projections. Therefore, we should note that many of the results in
Refs.~\cite{Mo02a,Mo02b,Mo02c,Mo02e,Mo02f,Mo02g,Mo03a,Mo04}
including the relation with $\PT$ symmetry, derived from the
assumption that there exists a complete set of
(bi)orthonormal eigenvectors, cannot be rigorously justified in
general, at least at present.

\section{Supersymmetry and Pseudo-supersymmetry}
\label{sec:sps}

Before discussing the relation between $\cN$-fold and nonlinear
pseudo-supersymmetries, we shall clarify in this section the simplest
case, namely, the relation between ordinary and
pseudo-supersymmetries. The Poincar\'e superalgebra in one spacetime
dimension is given by
\begin{align}
\label{eq:SUSY}
\bigl[\bA^{\pm},\bH\,\bigr]=0,\quad
 \bigl\{\bA^{\pm},\bA^{\pm}\bigr\}=0,\quad
 \bigl\{\bA^{-},\bA^{+}\bigr\}=2\bH.
\end{align}
An arbitrary system possessing the dynamical symmetry characterized by
the above superalgebra is given by a representation thereof. In
particular, a pair of Schr\"odinger operators $H^{\pm}$ can be
embedded into the following representation:
\begin{align}
\bA^{-}=\left(\begin{array}{cc}0&A^{-}\\0&0\end{array}\right),\quad
 \bA^{+}=\left(\begin{array}{cc}0&0\\A^{+}&0\end{array}\right),\quad
 \bH=\left(\begin{array}{cc}H^{+}&0\\0&H^{-}\end{array}\right)=
 \frac{1}{2}\left(\begin{array}{cc}A^{-}A^{+}&0\\
 0&A^{+}A^{-}\end{array}\right),
\label{eq:SS}
\end{align}
where the operators $A^{\pm}$ are given by
\begin{align}
A^{-}=\frac{\rmd}{\rmd x}+W(x),\qquad
 A^{+}=(A^{-})^{t}=-\frac{\rmd}{\rmd x}+W(x).
\end{align}
Arbitrary one-body supersymmetric quantum mechanical systems in
the literature are in fact mathematically equivalent to the above
system with a specific choice of the function $W(x)$, although various
notation, conventions, and terminology have been employed.

The crucial point here is that \emph{the superalgebra \eqref{eq:SUSY}
always holds for an arbitrary (differentiable) complex function
$W(x)$.} Then, if we restrict the function $W(x)$
to be real, $A^{+}$ is the formal Hermitian conjugate of $A^{-}$,
$A^{+}=(A^{-})^{\rmf}$, and the operators $H^{\pm}$ are formally
Hermitian, $(H^{\pm})^{\rmf}=H^{\pm}$. If we further restrict
the real function $W(x)$ to be in a special class of real functions,
it may be possible to define a Hilbert space $L^{2}(S)$
($S\subset\bbR$) on which $H^{\pm}$ are (rigorously) Hermitian,
$(H^{\pm})^{\rmf}=(H^{\pm})^{\dagger}=H^{\pm}$.
On the other hand, if we restrict $W(x)$ to a class of complex
functions such that there exists an invertible, formally Hermitian
operator $\eta$ for which the relation $\bA^{+}=\eta^{-1}(\bA^{-}
)^{\rmf}\eta$ holds, the operator $\bH$ is formally
pseudo-Hermitian, $\bH^{\rmf}= \eta\,\bH\,\eta^{-1}$. A further
restriction of the complex function $W(x)$ may enable us to define
a Hilbert space $L^{2}(S)$ on which $\bH$ is (rigorously)
pseudo-Hermitian \cite{Mo02a}. Finally, if the complex function
$W(x)$ satisfies $W^{\ast}(-x^{\ast})=-W(x)$, the operators $H^{\pm}$
are $\PT$-symmetric.

It is thus apparent that Hermitian, $\PT$-symmetric, or
pseudo-Hermitian supersymmetric systems are special cases of
\emph{general} supersymmetry, which is characterized by
the Poincar\'e superalgebra \eqref{eq:SUSY} in one spacetime
dimension, depending on the restrictions one imposes on the function
$W(x)$.

\section{$\cN$-fold and Nonlinear Pseudo-Supersymmetry}
\label{sec:Nsnps}

Next, we shall clarify the relation between $\cN$-fold and nonlinear
pseudo-supersymmetry. $\cN$-fold supersymmetry is characterized by
a superalgebra of the type
\begin{align}
\bigl[\bQ_{\cN}^{\pm},\bH_{\cN}\bigr]=0,\quad
 \bigl\{\bQ_{\cN}^{\pm},\bQ_{\cN}^{\pm}\bigr\}=0,\quad
 \bigl\{\bQ_{\cN}^{-},\bQ_{\cN}^{+}\bigr\}=\Pi_{\cN}(\bH_{\cN}),
\end{align}
where $\Pi_{\cN}$ is a polynomial of degree $\cN$. The operators
$\bQ_{\cN}^{\pm}$ are called $\cN$-fold supercharges. For a pair of
Schr\"odinger operators $H_{\cN}^{\pm}$ and a monic linear
differential operator $P_{\cN}$ of order $\cN$:
\begin{align}
P_{\cN}=\sum_{k=0}^{\cN}w_{k}(x)\frac{\rmd^{k}}{\rmd x^{k}},
\label{eq:Nfsc}
\end{align}
$\cN$-fold supersymmetry can be simply realized by the matrix
representation
\begin{align}
\bQ_{\cN}^{-}=\left(\begin{array}{cc}0&P_{\cN}\\
 0&0\end{array}\right),\quad
\bQ_{\cN}^{+}=\left(\begin{array}{cc}0&0\\
 P_{\cN}^{t}&0\end{array}\right),\quad
\bH_{\cN}=\left(\begin{array}{cc}H_{\cN}^{+}&0\\
 0&H_{\cN}^{-}\end{array}\right).
\label{eq:NS}
\end{align}
For a discussion of the general aspects of $\cN$-fold supersymmetry,
see, e.g., Refs.~\cite{AS03,AST01b,GT05}. In particular, the system
\eqref{eq:NS} reduces to the ordinary supersymmetric system
\eqref{eq:SS} when $\cN=1$.

From the discussion in the previous section, it should be almost
apparent that Hermitian, $\PT$-symmetric, or pseudo-Hermitian
$\cN$-fold supersymmetric systems can all be realized as special cases
of (general) $\cN$-fold supersymmetry, depending on the properties
of the functions $w_{k}(q)$ ($k=0,\dots,\cN$) in the component of
$\cN$-fold supercharge \eqref{eq:Nfsc}.

We now prove generically that any nonlinear pseudo-supersymmetric
system has $\cN$-fold supersymmetry. Indeed, since any nonlinear
pseudo-supersymmetric pair of differential operators $h_{0}$ and
$h_{N}$ (using the notation of Ref.~\cite{SR05}) satisfies, by
definition, intertwining relations with respect to higher-order
linear differential operators $A^{(N)}$ and $A^{(N)\natural}$:
\begin{align}
A^{(N)}h_{0}=h_{N}A^{(N)},\qquad
 h_{0}A^{(N)\natural}=A^{(N)\natural}h_{N},
\end{align}
the operators $h_{0}$ and $h_{N}$ preserve the finite-dimensional
vector spaces $\ker A^{(N)}$ and $\ker A^{(N)\natural}$, respectively,
and thus are weakly quasi-solvable. Applying the theorem on
the equivalence between weak quasi-solvability and $\cN$-fold
supersymmetry rigorously proved in Ref.~\cite{AST01b}, and using
the fact that the difference between $h_{0}$ and $h_{N}$ is uniquely
determined by the given $A^{(N)}$, we immediately conclude that
$h_{0}$ and $h_{N}$ must be an $\cN$-fold supersymmetric pair.\newpage

To illustrate the above fact more concretely, we shall show in what
follows how we can construct the nonlinear pseudo-supersymmetric
models in Ref.~\cite{SR05} in the framework of
$\cN$-fold supersymmetry with the aid of the algorithm developed
by us in Ref.~\cite{GT05}. Our starting point is the two-dimensional
linear space
\begin{align}
\label{eq:tv2}
\tcV_{2}=\bigl\langle 1,z\bigr\rangle,
\end{align}
and the most general linear second-order differential
operator preserving the latter space:
\begin{align}
\tH^{-}=-A(z)\frac{\rmd^{2}}{\rmd z^{2}}-B(z)\frac{\rmd}{\rmd z}
 -C(z),
\end{align}
where $A(z)$ is an arbitrary function, and $B(z)$ and $C(z)$ are
given by
\begin{align}
\label{eq:defB}
B(z)&=b_{2}z^{2}+b_{1}z+b_{0},\\
C(z)&=-b_{2}z+c_{0},
\end{align}
$b_{i}$ and $c_{0}$ being constants. Following the algorithm
for constructing an $\cN$-fold supersymmetric system developed in
Ref.~\cite{GT05}, we easily obtain the components of 2-fold
supersymmetry $(H^{\pm},P_{2})$ as follows:
\begin{align}
\label{eq:2Hampm}
H^{\pm}&=-\frac{1}{2}\frac{\rmd^{2}}{\rmd x^{2}}+\frac{1}{4A(z)}
 \biggl(\frac{A'(z)}{2}\pm B(z)\biggr)\biggl(\frac{3A'(z)}{2}\pm B(z)
 \biggr)-\frac{A''(z)}{4}\mp B'(z)-R,\\
\label{eq:2fsch}
P_{2}&=\frac{\rmd^{2}}{\rmd x^{2}}-\frac{2B(z)}{\dot{z}}
 \frac{\rmd}{\rmd x}-\frac{1}{2A(z)}\biggl(\frac{A'(z)}{2}-B(z)
 \biggr)\biggl(\frac{3A'(z)}{2}+B(z)\biggr)+\frac{A''(z)}{2}-B'(z),
\end{align}
where $R=b_{1}/2+c_{0}$, the dot and the prime denote derivative
with respect to $x$ and $z$, respectively, and the relation between
these two variables is determined by
\begin{align}
\label{eq:change}
\dot{z}^{2}=2A(z).
\end{align}
The solvable sector $\cV_{2}^{-}$ of the Hamiltonian $H^{-}$ is
given by
\begin{align}
\cV_{2}^{-}=\rme^{-\cW(z)}\bigl\langle 1,z \bigr\rangle,
\end{align}
with the gauge factor
\begin{align}
\label{eq:gauge}
\cW(z)=\int\frac{\rmd z}{2A(z)}\biggl(\frac{A'(z)}{2}-B(z)\biggr).
\end{align}

Let us first set the arbitrary function $A(z)$ as
\begin{align}
A(z)=8(2-a+z)\bigl(2-a+\sqrt{2-a+z}\bigr),
\end{align}
where $a$ is a parameter. From Eq.~\eqref{eq:change}, the change
of variable is given by
\begin{align}
z(x)=(2-a)(1-a)-2(2-a)\xb^{2}+\xb^{4},
\end{align}
where $\xb=x-x_{0}$, $x_{0}$ being a constant. Applying the
formulas \eqref{eq:2Hampm} and \eqref{eq:2fsch}, we obtain
the following 2-fold supersymmetric system $(H^{\pm},P_{2})$:
\begin{align}
H^{-}=&\,-\frac{1}{2}\frac{\rmd^{2}}{\rmd x^{2}}+b_{2}(\cdots)
 +\frac{b_{1}^{2}}{32}\xb^{2}+\frac{(2+(1-a)b_{1})(6+(1-a)b_{1})
 +2b_{0}b_{1}}{32\xb^{2}}\notag\\
&\,-\frac{48+8b_{1}-b_{1}^{2}}{32(2-a-\xb^{2})}
 +\frac{b_{0}(4-b_{1})}{16\xb^{2}(2-a-\xb^{2})}
 +\frac{b_{0}^{2}}{32\xb^{2}(2-a-\xb^{2})^{2}}\notag\\
&\,+\frac{(2-a)(48+16b_{1}+b_{1}^{2})-2b_{0}(8+b_{1})}{
 32(2-a-\xb^{2})^{2}}+\frac{b_{1}}{16}(4-2b_{1}+ab_{1})-R,\\
H^{+}=&\,-\frac{1}{2}\frac{\rmd^{2}}{\rmd x^{2}}+b_{2}(\cdots)
 +\frac{b_{1}^{2}}{32}\xb^{2}+\frac{(2-(1-a)b_{1})(6-(1-a)b_{1})
 +2b_{0}b_{1}}{32\xb^{2}}\notag\\
&\,-\frac{48-8b_{1}-b_{1}^{2}}{32(2-a-\xb^{2})}
 -\frac{b_{0}(4+b_{1})}{16\xb^{2}(2-a-\xb^{2})}
 +\frac{b_{0}^{2}}{32\xb^{2}(2-a-\xb^{2})^{2}}\notag\\
&\,+\frac{(2-a)(48-16b_{1}+b_{1}^{2})+2b_{0}(8-b_{1})}{
 32(2-a-\xb^{2})^{2}}-\frac{b_{1}}{16}(4+2b_{1}-ab_{1})-R,\\
P_{2}=&\,\frac{\rmd^{2}}{\rmd x^{2}}-\frac{1}{2}\biggl[ b_{2}(\cdots)
 +b_{1}\xb-\frac{(1-a)b_{1}}{\xb}+\frac{b_{1}\xb}{2-a-\xb^{2}}
 -\frac{b_{0}}{\xb(2-a-\xb^{2})}\biggr]\frac{\rmd}{\rmd x}\notag\\
&\,+b_{2}(\cdots)+\frac{b_{1}^{2}}{16}\xb^{2}-\frac{(2+(1-a)b_{1})
 (6-(1-a)b_{1})-2b_{0}b_{1}}{16\xb^{2}}\notag\\
&\,+\frac{48+4b_{1}+b_{1}^{2}}{16(2-a-\xb^{2})}
 -\frac{b_{0}(2+b_{1})}{8\xb^{2}(2-a-\xb^{2})}
 +\frac{b_{0}^{2}}{16\xb^{2}(2-a-\xb^{2})^{2}}\notag\\
&\,-\frac{(2-a)(48+8b_{1}-b_{1}^{2})-2b_{0}(4-b_{1})}{
 16(2-a-\xb^{2})^{2}}-\frac{b_{1}}{8}(2+2b_{1}-ab_{1}),
\end{align}
where each $b_{2}(\cdots)$ indicates a term proportional to $b_{2}$,
all of which are lengthy and thus will be abbreviated in this paper.
We easily see that the Hamiltonians $H^{\pm}$ are
$\PT$-symmetric, namely, invariant under the formal replacement
$x\to -x$, $\rmi\to -\rmi$, provided that the parameters are
chosen such that $a, b_{i},\rmi x_{0}\in\bbR$. Furthermore, one can
easily show that the above 2-fold supersymmetric system
$(2H^{-},2H^{+},P_{2}^{-}=P_{2},P_{2}^{+}=P_{2}^{t})$ exactly reduces
to the second nonlinear pseudo-supersymmetric system $(h_{0},h_{2},
A,A^{\natural})$ in Ref.~\cite{SR05}, Section 4.2, if we take the
parameters as $b_{2}=b_{0}=0$, $b_{1}=-4$ and $R=a-3$, with
$a=q\alpha$ and $x_{0}=\rmi\epsilon$.

Next, let us choose the function $A(z)$ as
\begin{align}
A(z)=\frac{1}{1-a}(8-4a-4z+z^{2})\Bigl[4-2a-3z+z^{2}-(1-z)
 \sqrt{8-4a-4z+z^{2}}\Bigr],
\end{align}
where the change of variable in this case is given by
\begin{align}
z(x)=\frac{(2-a)(1-a)-2(2-a)\xb^{2}+\xb^{4}}{1-a-\xb^{2}}.
\end{align}
Following the same procedure as in the previous case, we obtain
a 2-fold supersymmetric system which can be $\PT$-symmetric
and which exactly reduces to the first nonlinear
pseudo-supersymmetric system in Ref.~\cite{SR05}, Section 4.1,
when the parameters take the values $b_{2}=b_{0}=0$, $b_{1}=-2$
and $R=a-4$, with $a=q\alpha$ and $x_{0}=\rmi\epsilon$.

Similarly, if we take the function $A(z)$ as
\begin{align}
A(z)=-32(1-p-q)^{2}y(1-y)\bigl[3-4p-2(3-2p-2q)y\bigr]^{2},
\end{align}
where $y=\frac{1}{2}(1-\rmi\sinh x)$, the change of variable is
given by
\begin{align}
z(x)=(3-4p)(1-4p)-8(3-4p)(1-p-q)y+8(3-2p-2q)(1-p-q)y^{2}.
\end{align}
The 2-fold supersymmetric system in this case can be
$\PT$-symmetric with a proper choice of the parameters and
completely coincides with the third nonlinear pseudo-supersymmetric
model in Ref.~\cite{SR05}, Section 5, when $b_{2}=b_{0}=0$,
$b_{1}=2(1-p-q)$ and $R=\frac{1}{2}(2-2p-2q+p^{2}+2pq+q^{2})$.

Therefore, we have shown that all the nonlinear pseudo-supersymmetric
models in Ref.~\cite{SR05} can be constructed in the framework of
$\cN$-fold supersymmetry without any difficulty. More precisely,
note that the 2-fold supersymmetric system given by
\eqref{eq:2Hampm} and \eqref{eq:2fsch} is a realization of type A
2-fold supersymmetry\footnote{In this respect, we recall the
important fact that type A $\cN$-fold supersymmetry with $\cN=2$
is special due to the lack of the condition $\rmd^{5}A(z)/
\rmd z^{5}=0$ \cite{ANST01,Ta03a}. As a consequence, type A 2-fold
supersymmetric models are more general than the $\fsl(2)$
Lie-algebraic quasi-solvable models in Ref.~\cite{Tu88}.}.
The previous results thus imply that all the nonlinear
pseudo-supersymmetric models constructed in Ref.~\cite{SR05} belong
to type A 2-fold supersymmetry.

\section{Second-order {D}arboux Transformation and type {A}
 2-fold Supersymmetry}
\label{sec:2DtA2s}

We shall now prove the more general fact that an arbitrary one-body
quantum Hamiltonian which admits two (local) eigenfunctions in
closed form belongs to type A 2-fold supersymmetry, irrespective
of whether or not it is Hermitian, $\PT$-symmetric, pseudo-Hermitian,
and so on. Suppose, to this end, that the Hamiltonian $H$ under
consideration has two analytic solutions $\psi_{i}(x)$ and
$\psi_{j}(x)$ with some spectral parameters $\lambda_{i}$ and
$\lambda_{j}$, respectively:
\begin{align}
\label{eq:assume}
H\psi_{i}(x)=\lambda_{i}\psi_{i}(x),\qquad
 H\psi_{j}(x)=\lambda_{j}\psi_{j}(x).
\end{align}
We define two functions $z(x)$ and $\cW(z)$ by
\begin{align}
\label{eq:defzW}
z(x)=\frac{\psi_{j}(x)}{\psi_{i}(x)},\qquad
 \cW(z)=-\ln\psi_{i}(x).
\end{align}
Then, it is evident that the gauged Hamiltonian $\tH^{-}$ defined by
\begin{align}
\label{eq:defgH}
\tH^{-}=\rme^{\cW}H\rme^{-\cW}
\end{align}
preserves the vector space
\begin{align}
\label{eq:defv2}
\tcV_{2}=\langle 1,z\rangle.
\end{align}
Hence, we have a type A 2-fold supersymmetric system \eqref{eq:2Hampm}
and \eqref{eq:2fsch} if we follow the procedure described in the
previous section, with the specific choices of $z(x)$, $\cW(z)$
and $\tH^{-}$ given by Eqs.~\eqref{eq:defzW} and \eqref{eq:defgH}.
Therefore, all the models constructed from second-order Darboux
transformations with two exact solutions, including those in
Refs.~\cite{BS97,Sa99,FNN00,FMRS02,FMR03}, belong to type A 2-fold
supersymmetry. We note that we have not assumed whether or not
the original Hamiltonian $H$ is Hermitian, $\PT$-symmetric,
pseudo-Hermitian, and so on. In fact, with this procedure we can
obtain all the nonlinear pseudo-supersymmetric models in
Ref.~\cite{SR05}. Another point worth mentioning
is that the gauged Hamiltonian \eqref{eq:defgH} must be diagonal
in the basis \eqref{eq:defv2} because of the assumption
\eqref{eq:assume} and the choice \eqref{eq:defzW}. It follows
that the function $B(z)$ calculated from $A(z)=\dot{z}^{2}/2$
and $\cW(z)$ via the relation \eqref{eq:gauge} must be
proportional to $z$, which results in $b_{2}=b_{0}=0$ in
Eq.~\eqref{eq:defB}. This is the underlying reason why the nonlinear
pseudo-supersymmetric models of Ref.~\cite{SR05} always emerge when
$b_{2}=b_{0}=0$ in our previous arguments. This observation also
indicates that the framework of Darboux transformations of order
$\cN$ based on $\cN$ eigenfunctions is in general more restrictive
than the framework of $\cN$-fold supersymmetry, for arbitrary integer
$\cN >2$.

\section{Concluding Remarks}
\label{sec:concl}

One of the most important lessons drawn from the above results is the
recognition of the different characters of symmetries. The realization
of $\cN$-fold supersymmetry, including ordinary one, in terms of
linear differential operators is essentially \emph{local}, in
the sense that it is solely characterized by pointwise relations
through a superalgebra. That is exactly the reason why a couple of
significant aspects of $\cN$-fold supersymmetry has an intimate
relation with other local concepts such as quasi-solvability
\cite{AST01b} and transposition symmetry \cite{AS03}. It was
shown \cite{Ta06a} that the relation among these local concepts
is also crucial in another realization of $\cN$-fold supersymmetry
for von Roos operators \cite{vR83}. Higher-order Darboux
transformations also make sense at the local level.
On the other hand, the concepts of Hermiticity, pseudo-Hermiticity,
and so on are \emph{global}, in the sense that they make sense
rigorously only when they are formulated in a Hilbert space which
encodes global properties such as the domain of operators,
boundary conditions, and so on.

For a given Hilbert space, any (pseudo-)Hermitian operator
defined on it inevitably has a particular form. That is, any
(pseudo-)Hermitian linear differential operator defined on
$L^{2}(S)$ ($S\subset\bbR$) must be formally (pseudo-)Hermitian.
Hence, we can discuss whether or not $\cN$-fold supersymmetric
linear differential operators can be in addition formally
Hermitian, $\PT$-symmetric, or formally pseudo-Hermitian at
the local level without referring to a Hilbert space. If it is
the case, the system can possess both of these characteristic
features. For instance, a system which is $\cN$-fold
supersymmetric and formally Hermitian as well is weakly
quasi-solvable and, if there is a self-adjoint extension on
a suitable Hilbert space $L^{2}(S)$, its eigenvalues are all
real. What the authors of Ref.~\cite{SR05} have achieved is exactly
that they constructed a few 2-fold supersymmetric Schr\"odinger
operators which are $\PT$-symmetric as well. Needless to say,
this does not mean that they generalized the framework of
$\cN$-fold supersymmetry.

Regarding the relation between $\PT$ symmetry and
pseudo-Hermiticity, on the other hand, much more care must
be exercised. This is because an eigenvalue problem of
a $\PT$-symmetric operator is often defined on a complex
contour rather than on the real line. Due to this fact,
a $\PT$-symmetric linear differential operator which is
formally pseudo-Hermitian as well need not share the properties
of pseudo-Hermitian operators (provided that they are rigorously
justified) when the eigenvalue problem is set for it. In this
respect, there was an attempt to map $\PT$-symmetric eigenvalue
problems on a complex contour to those on the real line
\cite{Mo05a}. However, the method in Ref.~\cite{Mo05a} needs
the knowledge that the $\PT$ symmetry of the system is unbroken,
and thus would hardly apply in the general situation where we
cannot know \emph{a priori} whether or not $\PT$ symmetry is
dynamically broken.

\begin{acknowledgments}
This work was partially supported by Spain's DGI under the grant
No.~BFM2002-02646 (A. G.-L.) as well as by the National Science Council
of the Republic of China under the grant No.~NSC-93-2112-M-032-009
(T. T.).
\end{acknowledgments}



\bibliography{refsels}
\bibliographystyle{npb}



\end{document}